\documentclass[%
 reprint,
 amsmath,amssymb,
 aps,
]{revtex4-2}

\usepackage{graphicx}
\usepackage{dcolumn}
\usepackage{bm}

\usepackage{makecell}
\usepackage{wasysym}

\usepackage{hyperref}
\hypersetup{colorlinks=true, linkcolor=blue, citecolor=blue, filecolor=magenta, urlcolor=blue}

\begin{document}

\title{Symmetry Preservation in Commensurate Twisted Bilayers}

\author{David T. S. Perkins}
\email{d.perkins@lboro.ac.uk}
\affiliation{Department of Physics, Loughborough University, Loughborough LE11 3TU, England, United Kingdom}


\begin{abstract}
Symmetry plays a key role in materials hosting Dirac electrons and underpins our ability to completely flatten the Dirac cone through the tuning of physical parameters such as twisting in van der Waals heterostructures. The emergent moir\'{e} patterns in twisted bilayers at small twist angles appear, at first glance, to be independent of the initial stacking order and hence are only shifted when one layer is translated with respect to the other. However, when the twist angle is large, differences can be seen at the level of both the lattice and electronic structure in the case of twisted bilayer graphene. In this work, we first address the problem of twisted Kagome bilayers and show that the rotational and dihedral symmetry of high-symmetry Kagome bilayers is preserved for all commensurate twist angles with a 6-fold symmetric twist centre. Hence, we demonstrate that the exact symmetry of small twist angle systems depends upon the initial stacking of the bilayer. We further apply the principles of our method to twisted bilayer graphene with a 3-fold symmetric twist centre to recover the results of [E. J. Mele, \hyperlink{https://link.aps.org/doi/10.1103/PhysRevB.81.161405}{Phys. Rev. B \textbf{81}, 161405 (2010)}].
\end{abstract}

\maketitle

\section{Introduction}

Twisted bilayer graphene (TBG) is the forerunner for demonstrating how van der Waals (vdW) heterostructures can be tuned through the simple introduction of a rotational misalignment between layers. Since the discovery of superconductivity in TBG \cite{Cao2018}, there has been an intense effort to unveil the physics underpinning its emergent topological behaviour and strongly correlated phases \cite{Kang2019,Koshino2019,Lu2019,Lian2021,Song2022,Xie2021,Xie2023}. Prior to this though, TBG had been at the centre of several early theoretical works \cite{Lopes_dos_Santos2007,Shallcross2008,Mele2010,Bistritzer2010,Lopes_dos_Santos2012,Moon2012,Moon2013,Moon2014,Finocchiaro2017} and the observed magic angle had been predicted to coincide with the emergence of flat bands \cite{Bistritzer2011}. Inspired by the success of TBG, twisting in graphene on transition metal dichalcogenide heterostructures has now begun to be explored, particularly in the context of spintronics where twisting can be used to alter the proximity-induced couplings and spin texture of the graphene layer's Fermi surface to have drastic effects on spin-charge coupled transport \cite{Li2019,David2019,Peterfalvi2022,Veneri2022,Lee2022,Sun2023,Perkins2024,Milivojevic2024}. The goal of twisting is to engineer band structures with favourable characteristics that encourage sought after phases and responses.

The appearance of regions with a vanishing Fermi velocity in the electronic structure, such as flat bands and van Hove singularities (VHSs), allow for electron-electron interactions to drive a material's phase and yield divergences in the density of states (DOS). Of particular interest are higher-order VHSs (HOVHSs) which lead to a power-law divergence of the DOS \cite{Shtyk2017,Efremov2019,Chandrasekaran2023,Classen2024}. Evidence for tunability of VHSs to create HOVHSs has been observed in several materials including strontium ruthenates \cite{Tamai2008,Abarca_Morales2023} and graphene-based vdW heterostructures \cite{Rosenzweig2020,Seiler2024}. Theoretical works have also predicted their appearance in TBG \cite{Sherkunov2018,Yuan2019} and an exotic new phase called a supermetal due to the presence of a HOVHS \cite{Isobe2019,Castro2023}. It was shown in Ref. \cite{Chandrasekaran2020} that a system can be tuned to host a HOVHS if we have access to enough parameters (e.g. twisting) to adjust the underlying Hamiltonian. At a fundamental level, the types of HOVHS a system may host is set by its symmetry.

The Kagome lattice presents itself as promising system for studying the interplay of topology and strong correlations \cite{Bulut2005,Guo2009,Xu2015,Wang2023,Lin2024} due to its innately rich energy landscape exhibiting a completely flat band across the entire Brillouin zone (BZ), VHSs at the BZ edges, and Dirac electrons about the BZ corners \cite{Yin2022}. With the recent discovery of vanadium-based antimonides as a Kagome host \cite{Ortiz2019,Kang2022_CsVSb,Kim2023,Luo2023}, there has been intense research into other materials that might host both monolayer and bilayer Kagome configurations including Ti-based metals and Sn-based magnets \cite{Ye2018,Yin2020,Li2021,Li2023,Yang2024_Kagome}. Of particular note is the appearance of an effective Kagome monolayer without an atomic cage in metal-organic frameworks, where the Kagome lattice is formed of molecular orbitals in place of the usual atomic orbital lattices at the centre of most studies \cite{Fuchs2020}. As a two-dimensional (2D) system, the role of the Kagome lattice in vdW heterostructures and how its electronic structure can be further controlled is a natural next step to consider. Unlike graphene though, a Kagome bilayer has three intuitive high-symmetry stackings with different rotational symmetries \cite{Lima2019} -- $C_{6z}$, $C_{3z}$, and $C_{2z}$ -- with the first two mapping onto $AA$ and bernal stacked graphene, repsectively. Introducing a twist into these Kagome bilayers, tight binding models have predicted a flattening of the Dirac cone around the moir\'{e} BZ corners with magic angle around twice that of TBG, $\theta_{0} \simeq 2.28^{\circ}$ \cite{Lima2019}. This vanishing of the Fermi velocity via Dirac cone flattening is extremely sensitive to the symmetry of the system however \cite{Sheffer2023}, with the exact symmetry determining whether the magic angle corresponds to a significant reduction in the Fermi velocity or a truly zero value.

In this paper, we demonstrate that the rotational aspects of the point group symmetry for all high-symmetry Kagome bilayers will be preserved for any commensurate twist centred on a $C_{6z}$ symmetry point of an individual layer. Our approach is underpinned by an overlap function that measures that separation of rotational symmetry points and must vanish when they overlap. We use this general principle to reproduce the two classes of moir\'{e} patterns obtained for TBG to illustrate its wider applicability. We organise the paper as follows: section \ref{Setup_sec} starts by introducing the various high-symmetry stacking configurations for bilayer graphene and bilayer Kagome, provides a labeling scheme for commensurate twist angle, and explains how to identify the appearance of high-symmetry points (HSPs) in a moir\'{e} pattern. We then proceed to determine which rotational symmetry points (RSPs) are guaranteed to overlap and yield an HSP in section \ref{RSP_guaranteed_60deg_sec} and which RSP overlaps are avoided in section \ref{RSP_overlap_60deg_sec} when the twist is applied about the $C_{6z}$ symmetry point of one layer. Finally, in section \ref{RSP_overlaps_120deg_sec}, we recover the results of Mele \cite{Mele2010} for a twist applied about a $C_{3z}$ RSP of an individual layer and explain how this is connected to the $C_{6z}$ twist centre scenario. The symmetry analysis we provide here will act as a guide to understanding the emergence of HOVHSs in twisted Kagome systems. Moreover, it reveals how stacking order plays a central role in determining the allowed HOVHSs upon any commensurate twist and unveils a $D_{2}$ scenario that cannot arise upon twisting around a $C_{3z}$ symmetry site.

\section{Definitions and Setup} \label{Setup_sec}

\subsection{High-Symmetry Stacking Orders}

\begin{figure}
    \centering
    \includegraphics[width=\linewidth]{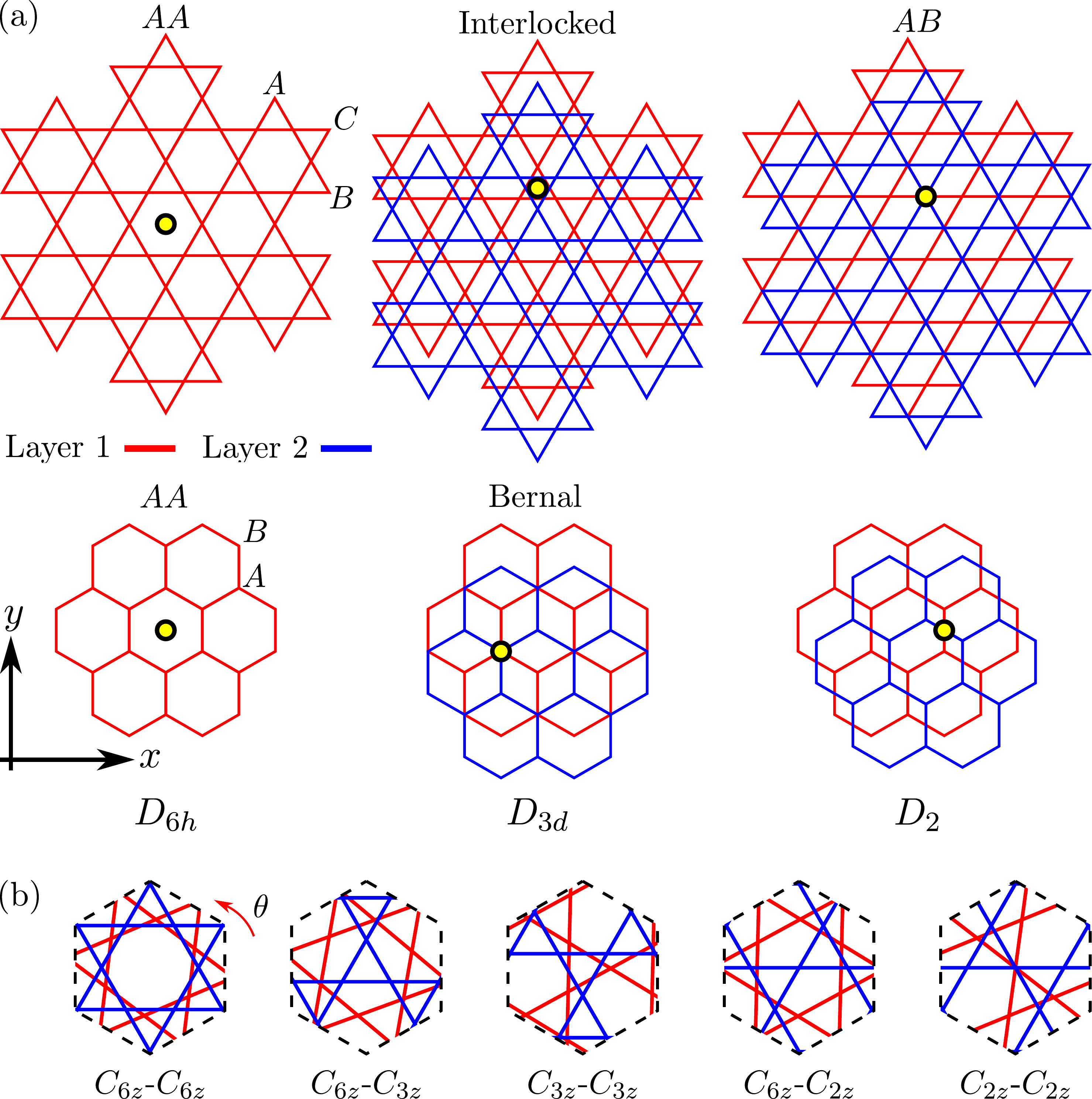}
    \caption{(a): High-symmetry stackings for Kagome bilayers (top) and their analogous graphene bilayer stacking (bottom). The yellow dots indicate high-symmetry centres. (b): Types of RSP overlap that yield a high-symmetry point.}
    \label{Stacking_orders}
\end{figure}

We define a high-symmetry stacking as one with a discrete rotational symmetry about the $z$-axis that is not identity, and that which possesses a $C_{2}$ symmetry about some in-plane axis (i.e. $C_{2}'$ symmetry). The resulting point group of the bilayer will thus be $D_{n}$, $D_{nv}$, or $D_{nh}$ depending on further reflection symmetries that may arise in the aligned bilayer. Fig. \ref{Stacking_orders}a presents all such high-symmetry stackings for graphene and Kagome bilayers prior to twisting. The stacking orders of bilayer graphene are already well documented in previous literature \cite{McCann2013}, so we spare a detailed discussion of its stacking order beyond stating that bilayer graphene has two high-symmetry stackings: $AA$ (column) with $D_{6h}$ symmetry and $AB$ (bernal) with $D_{3d}$ symmetry. In comparison, bilayer Kagome has only recently started to garner attention \cite{Ye2018,Lima2019,Yin2020,Li2021,Sinha2021}, and so let us take a moment to define the high-symmetry stackings possible in Kagome bilayers.

Due to hosting three sublattices, a Kagome bilayer also has three high-symmetry stackings. The most trivial to see is the $AA$ scenario, where one layer is placed directly on top of the other with no relative displacement. This scenario preserves all symmetries of the underlying Kagome monolayer and thus has a $D_{6h}$ point group symmetry. The HSP of the $AA$ bilayer Kagome is then located at the centre of the hexagon. We note that in all cases, the HSP will naturally have a vertical position halfway between the two layers and so we shall only discuss the 2D position of the bilayer HSP.

Turning our attention to the interlocked stacking of Kagome bilayers, we see that it is equivalent to bernal bilayer graphene, with the up and down triangles playing the role of graphene sublattices. To construct an interlocked Kagome bilayer, we place the second Kagome layer such that the centre of a down triangle from one layer lies directly on top of the centre of an up triangle from the other. Naturally, we see that the interlocked Kagome bilayer possesses the same point group symmetry as bernal bilayer graphene, $D_{3d}$, with the HSP being situated at the centre of an up-down triangle overlap.

Lastly, we may choose to arrange the Kagome bilayer in an $AB$ configuration where the second layer is shifted relative to the first along one of the bonds by a bond length. This results in two sublattices with direct overlap between the layers and one left floating in the centre of the hexagon from the other layer. In this case, whilst $C_{2z}$ and $C_{2}'$ are both present, no mirror planes of the constituent monolayers align meaning that $AB$ bilayer Kagome has no mirror reflection symmetries. Hence, the point group symmetry of this system is simply $D_{2}$. The HSP here is found where the sites of the two layers overlap one another directly.

\subsection{Twisting and Commensurate Angles} \label{Commensurate_angles_sec}

We introduce twisting into a high-symmetry bilayer by rotating one layer by an angle of $\theta$ about the $z$-axis and note that there are two natural centres that may be taken as the origin for this rotation. First, we may retain a $60^{\circ}$ twist periodicity, $\theta_{p} = 60^{\circ}$, for all choices of stacking by applying this rotation about the isolated HSP of the affected layer. Second, we may increase $\theta_{p}$ to that of the bilayer by choosing to rotate one layer about the HSP of the bilayer rather than of the individual layer. This will result in $\theta_{p} = 120^{\circ}$ for bernal TBG and interlocked twisted bilayer Kagome (TBK), whilst $\theta_{p} = 180^{\circ}$ will arise for $AB$ TBK. In this paper we shall consider the cases where $\theta_{p} = 60^{\circ}$ and $\theta_{p} = 120^{\circ}$.

The honeycomb and Kagome lattices are both examples of a hexagonal lattice with different sublattice geometries. Hence, their corresponding twisted bilayers possess the same set of commensurate twist angles, $\theta_{c}$. Each commensurate twist angle, $0 < \theta_{c} < 60^{\circ}$, is defined by a pair of coprime integers, $m$ and $n$, such that $m > n > 0$. The commensurate angle associated to $(m,n)$ is given by \cite{Shallcross2008,Shallcross2010,Scheer2022}
\begin{equation}
    \cos \theta_{c} = \frac{3m^{2} - n^{2}}{3m^{2} + n^{2}}, \quad \sin \theta_{c} = \frac{2\sqrt{3} \, mn}{3m^{2} + n^{2}}.
\end{equation}
For each $\theta_{c}$ there is a corresponding commensurate angle $\theta_{c}' = \pi/3 - \theta_{c}$. This pair of commensurate angles will always have one associated to when 3 divides $n$ and one where 3 does not divide $n$ \cite{Scheer2022}. When $n$ is divisible by 3, we choose to write $n = 3\nu$ ($\nu \in \mathbb{Z}$).

In the case where $\theta_{p} = 60^{\circ}$, we only need to consider one set of the these commensurate angles. Specifically, we choose to focus on those where $n$ is divisible by 3. This is a result of the layer being rotated possessing the mirror reflection symmetry $y \rightarrow -y$ about its HSP prior to rotation. We illustrate the equivalence of these lattices in the Appendix. For the cases where $\theta_{p} = 120^{\circ}$, we will need to consider both sets of $n$ as the commensurate angle pair $(\theta_{c},\theta_{c}')$ will yield different moir\'{e} patterns \cite{Mele2010}.

\begin{table}[t]
\caption{Resulting point group symmetry of a twisted bilayer due to the overlap of two RSPs (top row and leftmost column). Dihedral symmetry is only preserved when identical RSPs overlap. All other cases yield a point group that is a subgroup of the overlapping RSP point groups, with $E$ being the identity group. Visualisation of these overlaps are provided in Fig. \ref{Stacking_orders}b. \label{RSP_overlap_table}}
\begin{ruledtabular}
    \begin{tabular}{cccc}
        & $C_{6z}$ & $C_{3z}$ & $C_{2z}$
        \\
        \colrule
        $C_{6z}$ & $D_{6}$ & $C_{3z}$ & $C_{2z}$
        \\
        $C_{3z}$ & $C_{3z}$ & $D_{3}$ & $E$
        \\
        $C_{2z}$ & $C_{2z}$ & $E$ & $D_{2}$ \\
    \end{tabular}
\end{ruledtabular}
\end{table}

\subsection{The Role of High-Symmetry Overlap}

For a twisted bilayer to possess a non-trvial rotational symmetry, two points must overlap with compatible non-trivial rotational symmetries; we refer to these points of interest as rotational symmetry points (RSPs). Kagome and graphene possess three RSPs: $C_{6z}$ around the hexagon centres, $C_{3z}$ at the graphene sublattice sites and Kagome triangle centres, and $C_{2z}$ about the mid-point between graphene sublattices and at the sublattice sites of Kagome, see Fig. \ref{Stacking_orders}a. Let us start by considering $AA$ stacked bilayers. Naturally, when twisted to a commensurate angle, the $C_{6z}$ RSPs of each are guaranteed to overlap since we may identify the twist centre as a moir\'{e} lattice site with 6-fold symmetry. Therefore, the $AA$ bilayers will have $C_{6z}$ symmetry. In fact, they possess a $D_{6}$ point group symmetry since the mirror planes of the individual layers no longer align whilst a $C_{2}'$ symmetry persists.

\begin{figure}
    \centering
    \includegraphics[width=\linewidth]{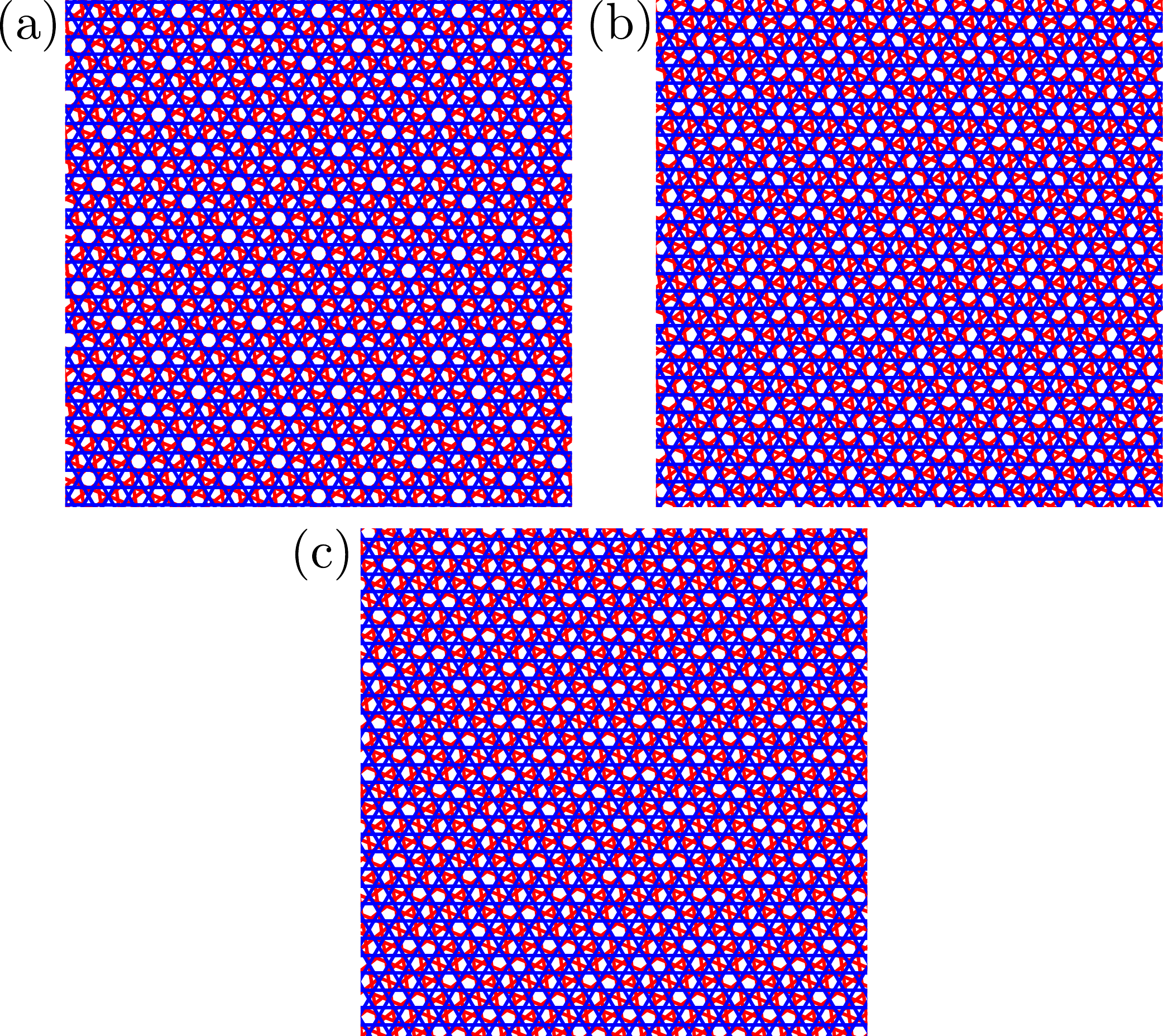}
    \caption{Moir\'{e} patterns for TBK in $AA$ (a), interlocked (b), and $AB$ (c) stacking. The twist is applied about a hexagon centre of the red layer to maintain $60^{\circ}$ twist periodicity. The commensurate angle used here is $\theta \simeq 38.21^{\circ}$ corresponding to $(m,n) = (5,3)$.}
    \label{Moire_patterns}
\end{figure}

If we now move on to consider the bernal/interlocked cases with a commensurate twist angle, we must study the emergent superlattice more closely as we cannot immediately say whether the twisted system will possess $C_{6z}$ or $C_{3z}$ symmetry. By inspection, we can see that $C_{3z}$ must comprise at least part, if not all, of the rotational symmetry group due to the twist centre always possessing this symmetry by construction for both $\theta_{p} = 60^{\circ}$ and $\theta_{p} = 120^{\circ}$, see Fig. \ref{Stacking_orders}b. Let us first focus on $\theta_{p} = 60^{\circ}$. Here, the twist centre corresponds to one case of RSP overlap, specifically when a $C_{6z}$ RSP overlaps a $C_{3z}$ RSP. However, two more cases must also be considered: $C_{3z}$ overlapping with $C_{3z}$, and $C_{6z}$ with $C_{6z}$. Each of these three cases yields a different point group symmetry which we summarise in Table \ref{RSP_overlap_table}. We must therefore check all RSP overlaps that may yield a higher point group symmetry. If we find that $C_{6z}$-$C_{6z}$ overlap is present, the system will host $D_{6}$ symmetry. However, if the system does not host $C_{6z}$-$C_{6z}$ overlap, we must then check for $C_{3z}$-$C_{3z}$ overlap which also preserves the dihedral symmetry of the original untwisted bilayers in contrast to the $C_{6z}$-$C_{3z}$ overlap naturally present in these systems. We find that, as shall be shown later, $C_{3z}$-$C_{3z}$ overlap does indeed occur for all commensurate twist angles in bernal TBG and interlocked TBK, whilst $C_{6z}$-$C_{6z}$ overlap will never occur with $\theta_{p} = 60^{\circ}$, and so the system will possess a $D_{3}$ point group symmetry, see Fig. \ref{Moire_patterns}b for an example with $\theta_{c} = 38.21^{\circ}$ ($m=5$ and $n=3$).

For the case where we instead apply our rotation about the bilayer HSP, $\theta_{p} = 120^{\circ}$, $C_{3z}$-$C_{3z}$ overlap occurs by construction, so we only need to check for $C_{6z}$-$C_{6z}$ overlap. Like the $\theta_{p} = 60^{\circ}$ case, every $\theta_{c}$ associated to when 3 divides $n$ avoids $C_{6z}$-$C_{6z}$ overlap to produce a $D_{3}$ point group symmetry. However, for $\theta_{c}$ with $n$ not divisible by 3, $C_{6z}$-$C_{6z}$ is guaranteed and the system instead exhibits $D_{6}$ symmetry \cite{Mele2010}.

Finally, for $AB$ stacked Kagome the twist centre corresponds to $C_{6z}$-$C_{2z}$ overlap, meaning the bilayer's minimum point group symmetry is $C_{2z}$. The overlaps we must check for are thus $C_{6z}$-$C_{6z}$, $C_{6z}$-$C_{3z}$, $C_{3z}$-$C_{3z}$, and $C_{2z}$-$C_{2z}$ as these will all yield higher point group symmetries than $C_{2z}$. Given that $C_{2z}$ and $C_{3z}$ are not compatible rotational symmetries, we do not have to check for their overlap. As we shall show later, we find that the only RSP overlap occurring besides $C_{6z}$-$C_{2z}$ for $\theta_{p} = 60^{\circ}$ is $C_{2z}$-$C_{2z}$, meaning that commensurate $AB$ TBK will possess a $D_{2}$ point group symmetry (Fig. \ref{Moire_patterns}c) when the twist origin is located at a hexagon centre.

\begin{table}[t]
\caption{Summary of the RSP locations within each layer for the various stacking choices for $\theta_{p} = 60^{\circ}$. We omit $AA$ stacking due to its trivial nature. The tildes on the subscripts for the $C_{3z}$ RSPs in $AB$ stacked Kagome indicate they are the lattice of up/down triangles analogous to the $A$ and $B$ sites of graphene. $\boldsymbol{\delta}_{\mathcal{O}} = \mathbf{0}$ indicates that the twist is centred on the $C_{6z}$ RSP of the bottom layer. \label{RSP_vectors}}
\begin{ruledtabular}
    \begin{tabular}{ccc}
        HSP & Bernal/Interlocked & $AB$ Kagome
        \\
        \colrule
        $C_{6}$ & \makecell{$\boldsymbol{\delta}_{\mathcal{O}}^{(\text{I})} = \mathbf{0}$, \\ $\boldsymbol{\delta}_{6'}^{(\text{I})} = \frac{\mathbf{a}_{1} + \mathbf{a}_{2}}{3}$} & \makecell{$\boldsymbol{\delta}_{\mathcal{O}}^{(\text{II})} = \mathbf{0}$, \\ $\boldsymbol{\delta}_{6'}^{(\text{II})} = \frac{\mathbf{a}_{2}-\mathbf{a}_{1}}{2}$}
        \\
        $C_{3}$ & \makecell{$\boldsymbol{\delta}_{A}^{(\text{I})} = \frac{2\mathbf{a}_{2}-\mathbf{a}_{1}}{3}$, \\ $\boldsymbol{\delta}_{B'}^{(\text{I})} = \frac{2\mathbf{a}_{2}-\mathbf{a}_{1}}{3}$} & \makecell{$\boldsymbol{\delta}_{\widetilde{A}}^{(\text{II})} = \frac{2\mathbf{a}_{2}-\mathbf{a}_{1}}{3}$, $\boldsymbol{\delta}_{\widetilde{B}}^{(\text{II})} = \frac{\mathbf{a}_{1} + \mathbf{a}_{2}}{3}$, \\ $\boldsymbol{\delta}_{\widetilde{A}'}^{(\text{II})} = \frac{\mathbf{a}_{1} + \mathbf{a}_{2}}{6}$, $\boldsymbol{\delta}_{\widetilde{B}'}^{(\text{II})} = -\frac{\mathbf{a}_{1} + \mathbf{a}_{2}}{6}$}
        \\
        $C_{2}$ & -- & \makecell{$\boldsymbol{\delta}_{A}^{(\text{II})} = \frac{\mathbf{a}_{1}}{2}$, $\boldsymbol{\delta}_{B}^{(\text{II})} = \frac{\mathbf{a}_{2}}{2}$, \\ $\boldsymbol{\delta}_{C}^{(\text{II})} = \frac{\mathbf{a}_{2}-\mathbf{a}_{1}}{2}$, $\boldsymbol{\delta}_{A'}^{(\text{II})} = \frac{2\mathbf{a}_{1}-\mathbf{a}_{2}}{2}$, \\ $\boldsymbol{\delta}_{B'}^{(\text{II})} = \frac{\mathbf{a}_{1}}{2}$, $\boldsymbol{\delta}_{C'}^{(\text{II})} = \mathbf{0}$} \\
    \end{tabular}
\end{ruledtabular}
\end{table}

\subsection{Identifying RSP Overlap} \label{Overlap_function_sec}

To determine the overlap of RSPs we construct a function that must vanish when two RSPs lie directly on top of one another. First, let $\boldsymbol{\delta}_{i}$ be the 2D position of the RSP of concern in the bottom layer and $\boldsymbol{\delta}_{j}$ be the 2D position of the RSP of concern in the top layer. The lattice of these RSPs is obtained by adding an arbitrary number of lattice vectors to their respective positions: $\mathbf{v}_{i} = \boldsymbol{\delta}_{i} + k \mathbf{a}_{1} + l \mathbf{a}_{2}$ and $\mathbf{v}_{j} = \boldsymbol{\delta}_{j} + p \mathbf{a}_{1} + q \mathbf{a}_{2}$, where $k,l,p,q \in \mathbb{Z}$. Next, we introduce the twist via the 2D rotation matrix $R_{\theta}$ such that $R_{\theta} \mathbf{v}_{i} = \mathbf{v}_{i}^{\theta}$. We now define our overlap function,
\begin{equation}
\begin{split}
    f(\boldsymbol{\delta}_{j},\boldsymbol{\delta}_{i};k,l,&p,q;\theta) \\
    &= \left| \boldsymbol{\delta}_{j} + p \mathbf{a}_{1} + q \mathbf{a}_{2} - \boldsymbol{\delta}_{i}^{\theta} - k \mathbf{a}_{1}^{\theta} - l \mathbf{a}_{2}^{\theta} \right|^{2}.
\end{split}
\end{equation}
Clearly, this vanishes when $\mathbf{v}_{i}^{\theta} = \mathbf{v}_{j}$. Specifically, if $f(\boldsymbol{\delta}_{j},\boldsymbol{\delta}_{i};k,l,p,q;\theta) = 0$ for some $k,l,p,q \in \mathbb{Z}$, then the RSPs at $\boldsymbol{\delta}_{i}$ and $\boldsymbol{\delta}_{j}$ must overlap at some point in the twisted lattice. We provide the full set of RSP positions to be checked in determining RSP overlap for the $\theta_{p} = 60^{\circ}$ in Table \ref{RSP_vectors}, where we denote the sublattice sites by $\{A,B,C\}$ ($\{A',B',C'\}$) for the bottom (top) layer. For $\theta_{p} = 120^{\circ}$, we need only check for $C_{6z}$-$C_{6z}$ overlap, and so only two RSP positions are required, $\widetilde{\boldsymbol{\delta}}_{6} = (\mathbf{a}_{1} - 2 \mathbf{a}_{2})/3 = - \widetilde{\boldsymbol{\delta}}_{6'}$ (taking the HSP as the origin), corresponding to the $C_{6z}$ RSP in the bottom and top layer, respectively.

A further property of the overlap function is that any zeros it exhibits must be minima since $f(\boldsymbol{\delta}_{j},\boldsymbol{\delta}_{i};k,l,p,q;\theta) \geq 0$. Therefore, a zero in the overlap function will always coincide with $\partial_{\mu} f(\boldsymbol{\delta}_{i},\boldsymbol{\delta}_{j};k,l,p,q;\theta) = 0$ for $\mu \in \{k,l,p,q\}$. These partial derivatives will allow us to establish additional relations between the lattice parameters $k$, $l$, $p$, and $q$ to determine whether they can be chosen to be integer such that the overlap function vanishes.

\section{Guaranteed RSP Overlaps: $\theta_{p} = 60^{\circ}$} \label{RSP_guaranteed_60deg_sec}

\subsection{Bernal TBG and Interlocked TBK}

We start by considering $C_{3z}$-$C_{3z}$ overlap in bernal TBG and interlocked TBK. We find that $\partial_{p} f(\boldsymbol{\delta}_{B'}^{(\text{I})},\boldsymbol{\delta}_{A}^{(\text{I})};k,l,p,q;\theta_{c}) = 0$ produces
\begin{equation}
    2p + q = \frac{(2k+l)m^{2} - 2(2+3l)m\nu - 3(2k+l)\nu^{2}}{m^{2}+3\nu^{3}}.
    \label{2pq_interlocked_full}
\end{equation}
In demanding that $p,q \in \mathbb{Z}$ we introduce a new parameter $\mathcal{A} = 2p + q$, $\mathcal{A} \in \mathbb{Z}$. Furthermore, by letting $l = -2k$,
\begin{equation}
    \mathcal{A} = \frac{4(3k-1)m\nu}{m^{2}+3\nu^{3}}.
    \label{Ak_relation_interlocked_simple}
\end{equation}
We next notice that $m^{2}+3\nu^{3}-1$ will always be divisible by 3 courtesy of the coprime nature of $m$ and $n$ such that $n$ is divisible by 3. To see this, we may write $m = 3\mu \pm 1$ ($\mu \in \mathbb{Z}$) to acquire $m^{2}+3\nu^{3}-1 = 3(3\mu^{2} \pm 2\mu + \nu^{2})$. We therefore choose
\begin{equation}
    k = \frac{2 + \alpha(m^{2} + 3 \nu^{2})}{3} (m^{2}+3\nu^{2}-1) + 1, \quad (\alpha \in \mathbb{Z}),
    \label{k_interlocked_choice}
\end{equation}
to ensure that both $\mathcal{A},l \in \mathbb{Z}$,
\begin{equation}
    \mathcal{A} = 4m\nu (2 + \alpha(m^{2}+3\nu^{2}-1)).
    \label{A_interlocked_integer_form}
\end{equation}

\begin{widetext}
Next, let us consider the overlap function with $k$ given by eq. \ref{k_interlocked_choice}, $l = -2k$, $q = \mathcal{A} - 2p$, and $\mathcal{A}$ given by eq. \ref{A_interlocked_integer_form},
\begin{equation}
    f(\boldsymbol{\delta}_{B'}^{(\text{I})},\boldsymbol{\delta}_{A}^{(\text{I})};k,l,p,q;\theta_{c}) =\frac{1}{3} \Big[ 1 - 3p + (m^{2}+6m\nu-3\nu^{2}) (2+\alpha(m^{2}+3\nu^{2}-1)) \Big]^{2}.
\end{equation}
The overlap function will thus vanish when
\begin{equation}
    p = \frac{1}{3} \Big[ 1 + (m^{2}+6m\nu-3\nu^{2}) (2+\alpha(m^{2}+3\nu^{2}-1)) \Big],
\end{equation}
which we see to be integer upon substituting $m = 3\mu \pm 1$ to obtain a cumbersome but nonetheless interger expression for $p$,
\begin{equation}
    p = 1 + \mu (3\mu \pm 2) (2 + \alpha (1 \pm 3 \mu)^{2}) + 2 [6(1+\alpha)\mu + 27\alpha \mu^{2} (\mu \pm 1) \pm 2] \nu + (\alpha - 2) \nu^{2} + 6 \alpha (3 \mu \pm 1) \nu^{3} - 3 \alpha \nu^{4}.
    \label{Interlocked_p_int_form}
\end{equation}
\end{widetext}

Therefore, $C_{3z}$-$C_{3z}$ overlap is guaranteed to occur in bernal TBG and interlocked TBK at all commensurate twist angles when the twist is applied about a $C_{6z}$ symmetry point.

\subsection{$AB$ TBK} \label{AB_TBK_guaranteed_RSP_overlap_sec}

In the commensurate $AB$ TBK system, we shall consider the overlap of $C_{2z}$ RSPs. Naturally, there are several combinations we might consider: $A$-$A'$, $A$-$B'$, $A$-$C'$, $B$-$C'$, and their layer flipped counterparts. We shall choose to focus on $C$-$A'$, $A$-$B'$, and $A$-$A'$ overlap depending on our choice of $m$ and $n$,
\begin{equation}
\begin{split}
    C\text{-}A': &\, m + \nu \text{ is even s.t. } \frac{m + \nu + 2}{4} \in \mathbb{Z},
    \\
    A\text{-}B': &\, m + \nu \text{ is even s.t. } \frac{m + \nu}{4} \in \mathbb{Z},
    \\
    A\text{-}A': &\, m + \nu \text{ is odd}.
\end{split}
\end{equation}
We justify these categorisations in the next section.

\subsubsection{$C$-$A'$ Overlap}

Let us move to the twist-symmetric frame wherein the bottom layer is rotated by $\theta_{c}/2$ and the top layer is rotated by $-\theta_{c}/2$. Clearly, this system is identical to the previous scenario in which only the bottom layer is rotated by $\theta_{c}$ through a global rotation of $-\theta_{c}/2$. The $C$ and $A'$ sites are then given by
\begin{subequations}
\begin{align}
    \mathbf{v}_{C} &= R_{\frac{\theta_{c}}{2}} (\boldsymbol{\delta}_{C}^{\text{(II)}} + k \mathbf{a}_{1} + l \mathbf{a}_{2}),
    \\
    \mathbf{v}_{A'} &= R_{-\frac{\theta_{c}}{2}} (\boldsymbol{\delta}_{A'}^{\text{(II)}} + p \mathbf{a}_{1} + q \mathbf{a}_{2}),
\end{align}
\end{subequations}
respectively. In this frame we now search for $C$-$A'$ overlap along the $x$-axis. Setting the $y$-components $\mathbf{v}_{C}$ and $\mathbf{v}_{A'}$ to zero yields
\begin{subequations}
\begin{align}
    4\nu k + 2(m+\nu) l &= -(m-\nu),
    \label{CAp_kl_Bezouts_equation}
    \\
    4\nu p - 2(m-\nu) q &= -(m+3\nu).
    \label{CAp_pq_Bezouts_equation}
\end{align}
\label{CAp_Bezouts_equations}%
\end{subequations}
The existence of integer solutions to these diophantine equations can be checked via B\'{e}zout's identity \cite{Coppel_book}.

\textit{B\'{e}zout's Identity:} If $a,b \in \mathbb{Z}$ and $d = \alpha \, \text{GCD}(a,b)$ with $\alpha \in \mathbb{Z}$ (i.e. $d$ is an integer multiple of the greatest common divisor of $a$ and $b$), then $\exists  \, x,y \in \mathbb{Z}$ such that $ax + by = d$.

Starting with eq. \ref{CAp_kl_Bezouts_equation}, we immediately note that when either $m$ or $\nu$ is even and the other is odd (both cannot be even due to $m$ and $3\nu$ being coprime) then $m \pm \nu$ must be odd and $\text{GCD}(4\nu,2(m+\nu)) = 2$. Hence, eq. \ref{CAp_kl_Bezouts_equation} cannot satisfy B\'{e}zout's identity when $m$ and $\nu$ possess opposite parity. The same can be said for eq. \ref{CAp_pq_Bezouts_equation} since $(m+3\nu)$ will also be odd when $m$ and $\nu$ are of opposite parity.

To progress, we consider the case when $m$ and $\nu$ are both odd, hence $m \pm \nu$ is even, and further divide this set of $m$ and $\nu$ into two equal sets: (i) those where $(m + \nu)/4 \in \mathbb{Z}$ and (ii) those where $(m + \nu)/4 \notin \mathbb{Z}$. In identifying these sets, we note that when 4 divides $(m+\nu)$, $m-\nu$ is even but not divisible by 4 and vice versa. Focusing on (i), we see that $\text{GCD}(4\nu,2(m \pm \nu)) = 4$. However, 4 does not divide $(m - \nu)$ and $(m+3\nu)$ here: letting $m + \nu = 4\alpha$ ($\alpha \in \mathbb{Z}$) we find $m - \nu = 2(2\alpha - \nu)$ and $m + 3\nu = 2(2\alpha + \nu)$. Therefore, eqs. \ref{CAp_kl_Bezouts_equation} and \ref{CAp_pq_Bezouts_equation} again do not satisfy B\'{e}zout's identity in this scenario. Moving onto (ii), we again find $\text{GCD}(4\nu,2(m \pm \nu)) = 4$ but now 4 divides $(m - \nu)$ and $(m+3\nu)$: letting $m + \nu + 2 = 4\alpha$ yields $m - \nu = 2(2\alpha - \nu - 1)$ and $m + 3\nu = 2(2\alpha + \nu - 1)$ and we note that $\nu \pm 1$ is even. Hence, the expressions in eq. \ref{CAp_Bezouts_equations} satisfy B\'{e}zout's identity when $m$ and $\nu$ are both odd such that $(m+\nu+2)$ is divisible by 4.

Having identified $\text{GCD}((4,m+\nu),2) = 2$ with $m$ and $\nu$ being odd as the relevant scenario, we use the following ansatz to ensure that $k,p \in \mathbb{Z}$,
\begin{equation}
    k = \frac{m+\nu+2}{4}, \qquad p = \frac{m-\nu}{4} - 1.
    \label{Scenario2_integers1}
\end{equation}
Substituting these expressions for $k$ and $p$ into eq. \ref{CAp_Bezouts_equations} yields
\begin{equation}
    l = -\frac{\nu+1}{2}, \qquad q = \frac{1+\nu}{2},
    \label{Scenario2_integers2}
\end{equation}
which are also integer given that $\nu$ is odd. Thus, eqs. \ref{Scenario2_integers1} and \ref{Scenario2_integers2} correspond to when $C$ and $A'$ sites lie on the $x$-axis in the twist-symmetric frame, respectively. Upon substitution of these expressions into the overlap function, we find that $f(\boldsymbol{\delta}_{C}^{\text{(II)}},\boldsymbol{\delta}_{A'}^{\text{(II)}};k,l,p,q;\theta_{c}) = 0$ demonstrating that the $C$ and $A'$ sites possessing $C_{2z}$ symmetry are guaranteed to overlap in scenario (ii) when $m$ and $\nu$ are both odd, leading to a system with at least $D_{2}$ point group symmetry.

\subsubsection{$A$-$B'$ Overlap}

We again consider the twist-symmetric frame but now focus on the apperance of $A$ and $B'$ sites along the $x$-axis in this frame,
\begin{subequations}
\begin{align}
    \mathbf{v}_{A} &= R_{\frac{\theta_{c}}{2}} (\boldsymbol{\delta}_{A}^{(\text{II})} + k \mathbf{a}_{1} + l \mathbf{a}_{2}),
    \\
    \mathbf{v}_{B'} &= R_{-\frac{\theta_{c}}{2}} (\boldsymbol{\delta}_{B'}^{(\text{II})} + p \mathbf{a}_{1} + q \mathbf{a}_{2}).
\end{align}
\end{subequations}
Demanding that the $y$-component of these vectors vanishes results in the following diophantine equations for the lattice parameters,
\begin{subequations}
\begin{align}
    2\nu k + (m+\nu) l &= -\nu,
    \label{ABp_kl_Bezouts_equation}
    \\
    2\nu p - (m-\nu) q &= -\nu.
    \label{ABp_pq_Bezouts_equation}
\end{align}
\label{ABp_Bezouts_equations}%
\end{subequations}
In this case, we find that only when $m$ and $\nu$ are of opposite parity can the expressions in eq. \ref{ABp_Bezouts_equations} satisfy B\'{e}zout's identity,
\begin{equation}
\begin{split}
    \text{GCD}(2\nu,(m\pm\nu)) = \begin{cases}
        1, \quad \mathcal{P}_{m} \neq \mathcal{P}_{\nu} \\
        2, \quad \mathcal{P}_{m} = \mathcal{P}_{\nu} = -1 
    \end{cases},
\end{split}
\end{equation}
where $\mathcal{P}_{m} = \pm 1$ when $m$ is even/odd (i.e. $\mathcal{P}_{m}$ is the parity of $m$). The latter case requires $\nu$ to be odd and hence not divisible by $\text{GCD}(2\nu,(m\pm\nu))$, where as the former case is always satisfied for any integer choice of $\nu$.

With the above in mind, we employ the ansatz
\begin{equation}
    k =\frac{m + \nu - 1}{2}, \qquad p = \frac{m - \nu - 1}{2},
    \label{Scenario0_integers1}
\end{equation}
which are guaranteed to be integer when $\mathcal{P}_{m} \neq \mathcal{P}_{\nu}$, and substitute eq. \ref{Scenario0_integers1} into eq. \ref{ABp_Bezouts_equations} to arrive at
\begin{equation}
    l = -\nu, \qquad q = \nu.
    \label{Scenario0_integers2}
\end{equation}
Using the integer expressions in eqs. \ref{Scenario0_integers1} and \ref{Scenario0_integers2} we find that the overlap function for the $A$ and $B'$ vanishes, $f(\boldsymbol{\delta}_{A}^{\text{(II)}},\boldsymbol{\delta}_{B'}^{\text{(II)}};k,l,p,q;\theta_{c}) = 0$. Therefore, $A$ sites will always overlap with $B$ sites when $m$ and $\nu$ are opposite in parity, again resulting in a system with at least $D_{2}$ symmetry.

\subsubsection{$A$-$A'$ Overlap}

Moving once more into the twist-symmetric frame, we apply an additional global rotation of $-\pi/6$ and then search for overlap of $A$ and $A'$ sites along the $x$-axis. In this frame of reference, the positions of the $A$ and $A'$ sites are given by
\begin{subequations}
\begin{align}
    \widetilde{\mathbf{v}}_{A} &= R_{\frac{\theta_{c}}{2}-\frac{\pi}{6}} (\boldsymbol{\delta}_{A}^{\text{(II)}} + k \mathbf{a}_{1} + l \mathbf{a}_{2}),
    \\
    \widetilde{\mathbf{v}}_{A'} &= R_{-\frac{\theta_{c}}{2}-\frac{\pi}{6}} (\boldsymbol{\delta}_{A'}^{\text{(II)}} + p \mathbf{a}_{1} + q \mathbf{a}_{2}).
\end{align}
\end{subequations}
Setting the $y$ components of these vectors to zero, we obtain
\begin{subequations}
\begin{align}
    2(m-3\nu) k - 2(m+3\nu) l &= -(m-3\nu),
    \label{AAp_kl_Bezouts_equation}
    \\
    2(m+3\nu) p - 2(m-3\nu) q &= -3(m+\nu).
    \label{AAp_pq_Bezouts_equation}
\end{align}
\label{AAp_Bezouts_equations}%
\end{subequations}
In checking B\'{e}zout's identity, we note that
\begin{equation}
\begin{split}
    &\text{GCD}(2(m-3\nu),2(m+3\nu)) \\
    &\qquad\qquad\qquad\qquad= \begin{cases}
        2, \quad \mathcal{P}_{m} \neq \mathcal{P}_{\nu}  \\
        4, \quad \mathcal{P}_{m} = \mathcal{P}_{\nu} = -1
    \end{cases},
\end{split}
\end{equation}
and hence the diophantine equations in eq. \ref{AAp_Bezouts_equations} can only satisfy B\'{e}zout's identity if 4 divides $(m+\nu)$, which in turn gives 4 divides $(m-3\nu)$. We thus see that the current setup corresponds to (i) when $m$ and $\nu$ are both odd.

We next use the ansatz
\begin{equation}
    k = \frac{m + 3\nu - 2}{4}, \qquad p = \frac{m - 3\nu}{4} - 1,
    \label{Scenario1_integers1}
\end{equation}
noting that when $(m + \nu)$ is divisible by 4 whilst $m$ and $\nu$ are both odd, 4 does not divide $(m + 3\nu)$ despite $m + 3\nu$ being even, implying that 4 divides $(m + 3\nu \pm 2)$. By substituting eq. \ref{Scenario1_integers1} into eq. \ref{AAp_Bezouts_equations}, we find
\begin{equation}
    l = \frac{m - 3\nu}{4}, \qquad q = \frac{m + 3\nu +2}{4}.
    \label{Scenario1_integers2}
\end{equation}
Finally, by considering the overlap function for sites $A$ and $A'$ with the lattice parameters identified in eqs. \ref{Scenario1_integers1} and \ref{Scenario1_integers2}, we find a vanishing result, $f(\boldsymbol{\delta}_{A}^{\text{(II)}},\boldsymbol{\delta}_{A'}^{\text{(II)}};k,l,p,q;\theta_{c}) = 0$. Therefore, in the final case where $m$ and $\nu$ are both odd such that their sum is divisible by 4, $A$ and $A'$ sites are guaranteed to overlap and hence the system possesses at least a $D_{2}$ point group symmetry.

\section{Avoided RSP Overlaps: $\theta_{p} = 60^{\circ}$} \label{RSP_overlap_60deg_sec}

Having demonstrated that $C_{jz}$-$C_{jz}$ ($j=2,3$) is guaranteed in commensurate twisted graphene and Kagome bilayers that started in a high-symmetry stacking configuration, with the rotational symmetry matching that of the untwisted system, we now move on to proving that overlap of RSPs yielding a higher rotational symmetry cannot occur. This section shall draw solely on searching for the minimum of the overlap function as explained in section \ref{Overlap_function_sec}.

\subsection{Bernal TBG and Interlocked TBK}

Starting with the bernal/interlocked systems, we find that $\partial_{p} f(\boldsymbol{\delta}_{6'}^{(\text{I})},\boldsymbol{\delta}_{\mathcal{O}}^{(\text{I})};k,l,p,q;\theta_{c}) = 0$ imposes
\begin{equation}
\begin{split}
    \mathcal{A} &= 2p + q \\
    &= \frac{(2k+l-1) m^{2} - 6 l m\nu - 3 (2k+l+1) \nu^{2}}{m^{2} + 3 \nu^{2}}.
\end{split}
\end{equation}
Rewritting this, we arrive at an expression for $k$ in terms of $\mathcal{A}$ and $l$,
\begin{equation}
    k = \frac{(1+\mathcal{A}-l) m^{2} + 6 l m \nu + 3 (1+\mathcal{A}+l) \nu^{2}}{2 (m^{2} - 3 \nu^{2})}.
    \label{k_C6C6_overlap_interlocked}
\end{equation}
Let us now assume that $\exists \, l,\mathcal{A} \in \mathbb{Z} \text{ s.t. } k \in \mathbb{Z}$. Substituting eq. \ref{k_C6C6_overlap_interlocked} and $q = \mathcal{A} - 2p$ into the overlap function $f(\boldsymbol{\delta}_{6'}^{(\text{I})},\boldsymbol{\delta}_{\mathcal{O}}^{(\text{I})};k,l,p,q;\theta_{c}) = 0$ yields
\begin{equation}
\begin{split}
    l = \frac{1}{m^{2} + 3 \nu^{2}} \bigg[ m^{2} (\mathcal{A} &- 2 p) - 
   2 (1 + A) m \nu \\
   &- (1 + 3 \mathcal{A} - 6 p) \nu^2 + \frac{m^{2}}{3} \bigg].
\end{split}
\end{equation}
Since $m$ cannot be divided by 3, we immediately see that $l$ cannot be integer when the overlap function vanishes. Therefore, by contradiction, the $C_{6}$ RSPs in the bernal/interlocked cases never overlap. The point group of bernal TBG and interlocked TBK is thus $D_{3}$.

\subsection{$AB$ TBK}

The case of $AB$ TBK requires more steps to establish that the highest symmetry RSP overlap is $C_{2z}$. In addition to checking $C_{6z}$-$C_{6z}$ overlap, we must check all possible combinations of $C_{6z}$-$C_{3z}$ and $C_{3z}$-$C_{3z}$ overlap.

\subsubsection{$C_{6z}$-$C_{6z}$ Overlap} \label{AB_C6_avoided_sec}

Starting with the 6-fold overlap, we find that $\partial_{p} f(\boldsymbol{\delta}_{6'}^{(\text{II})},\boldsymbol{\delta}_{\mathcal{O}}^{(\text{II})};k,l,p,q;\theta_{c}) = 0$ yields
\begin{equation}
    \mathcal{A}_{\hexagon}^{\hexagon} = \frac{1}{2} - 2k - l - \frac{2m[2m k + (m-3\nu)l]}{m^{2} + 3 \nu^{2}},
    \label{A_C6C6_overlap_interlocked}
\end{equation}
where we use the superscript and subscript on $\mathcal{A}$ to denote the RSP symmetry of the top/bottom layer in the overlap function, respectively. Let us consider when $m$ and $\nu$ are of opposite parity. In this case, $m^{2} + 3 \nu^{2}$ will be odd and the final term of eq. \ref{A_C6C6_overlap_interlocked} will never yield a result to cancel the factor of $1/2$. Turning our attention to when $m$ and $\nu$ are both odd, we note that $m^{2} + 3 \nu^{2}$ will now be even at the same time as $2m$ and $m-3\nu$ becoming even. Then, for the final term of eq. \ref{A_C6C6_overlap_interlocked} to be a half integer to counteract the leading factor of $1/2$, $m^{2} + 3 \nu^{2}$ will need to be divisible by $2^{\gamma}$ with $\gamma \in \mathbb{Z}$ and $\gamma \geq 3$. However, this cannot happen when $m$ and $\nu$ are both odd: letting $m = 2\alpha + 1$ and $\nu = 2\beta + 1$ $(\alpha,\beta \in \mathbb{Z})$ gives $m^{2} + 3 \nu^{2} = 4[1 + \alpha(1+\alpha) + 3\beta(1+\beta)]$. Consequently, it is not possible for a minimum of the overlap function to exist in $p$ that corrseponds to an integer value of $\mathcal{A}$, which in turn means that there are no integer choices for $p$ and $q$ corresponding to a vanishing of the overlap function. Therefore, $C_{6z}$-$C_{6z}$ overlap in commensurate $AB$ TBK does not occur.

\subsubsection{$C_{6z}$-$C_{3z}$ Overlap}

There are four cases that lead to $C_{6z}$-$C_{3z}$ overlap (Ia): an up triangle of the top layer with a hexagon centre of the bottom layer, (IIa): a down triangle of the top layer with a hexagon centre of the bottom layer, (IIIa): an up triangle of the bottom layer with a hexagon centre of the top layer, (IVa): a down triangle of the bottom layer with a hexagon centre of the top layer. By considering $\partial_{p} f(\boldsymbol{\delta}_{j}^{(\text{II})},\boldsymbol{\delta}_{i}^{(\text{II})};k,l,p,q;\theta_{c}) = 0$ for cases (Ia) and (IIa), we obtain expressions for $\mathcal{A}_{\hexagon}^{\vartriangle}$ and $\mathcal{A}_{\hexagon}^{\triangledown}$ that are related to $\mathcal{A}_{\hexagon}^{\hexagon}$ by an integer shift,
\begin{subequations}
\begin{align}
    \text{(Ia):}& \quad (i,j) = (\mathcal{O},\widetilde{A}') \quad \Rightarrow \quad \mathcal{A}_{\hexagon}^{\vartriangle} = \mathcal{A}_{\hexagon}^{\hexagon} - 1,
    \\
    \text{(IIa):}& \quad (i,j) = (\mathcal{O},\widetilde{B}') \quad \Rightarrow \quad \mathcal{A}_{\hexagon}^{\triangledown} = \mathcal{A}_{\hexagon}^{\hexagon}.
\end{align}
\end{subequations}
Hence, the same reasoning applies here as in section \ref{AB_C6_avoided_sec}. Next, we consider the layer reversed situation with the $C_{6z}$ RSP located in the top layer and $C_{3z}$ RSP in the bottom layer, wherein
\begin{subequations}
\begin{align}
    \mathcal{A}_{\vartriangle}^{\hexagon} &= \frac{1}{2} - 2k - l + \frac{2m[2m k + (m-3\nu)l - 2\nu]}{m^{2} + 3 \nu^{2}}
    \label{hex_up_A}
    \\
    \mathcal{A}_{\triangledown}^{\hexagon} &= \frac{1}{2} - 2k - \tilde{l} + \frac{2m[2m k + (m-3\nu)\tilde{l} + 2\nu]}{m^{2} + 3 \nu^{2}}
\end{align}
\end{subequations}
for cases (IIIa) and (IVa), respectively, where $\tilde{l} = l - 1$. The fourth term of $\mathcal{A}_{\vartriangle,\triangledown}^{\hexagon}$ again fails to compensate for the leading term of 1/2 by the same reasoning as used for $\mathcal{A}_{\hexagon}^{\hexagon}$. To see this, we note that only when $m$ and $\nu$ are both odd is the denominator of the fourth term even, and so letting $m = 2\alpha + 1$ and $\nu = 2\beta + 1$ $(\alpha,\beta \in \mathbb{Z})$ yields
\begin{equation}
    (1+2\alpha) \frac{1 + (1+2\alpha)k + (\alpha-3\beta-1)l - 2\beta}{1 + \alpha(1+\alpha) + 3\beta(1+\beta)}
\end{equation}
for the final term of eq. \ref{hex_up_A}. The denominator of this writing of the final term will always be odd and hence cannot enable the 1/2 to be cancelled in $\mathcal{A}_{\vartriangle}^{\hexagon}$.

\subsubsection{$C_{3z}$-$C_{3z}$ Overlap}

We again have four cases that lead to $C_{3z}$-$C_{3z}$ overlap (Ib): up triangle overlap, (IIb): down triangle overlap, (IIIb): an up triangle of the top layer with a down triangle of the bottom layer, (IVb): a down triangle of the top layer with an up triangle of the bottom layer. The avoidance of RSP in each of these cases collapses to the same reasoning, so let us consider (Ib) as the main example for $C_{3z}$-$C_{3z}$ overlap. Once again, in setting $\partial_{p} f(\boldsymbol{\delta}_{j}^{(\text{II})},\boldsymbol{\delta}_{i}^{(\text{II})};k,l,p,q;\theta_{c}) = 0$ for each of these cases, we obtain expressions for the various $\mathcal{A}_{\vartriangle,\triangledown}^{\vartriangle,\triangledown}$ that are related to $\mathcal{A}_{\vartriangle,\triangledown}^{\hexagon}$ by simple integer shifts,
\begin{subequations}
\begin{align}
    \text{(Ib):}& \quad (i,j) = (\widetilde{A},\widetilde{A}') \,\, \Rightarrow \,\, \mathcal{A}_{\vartriangle}^{\vartriangle} = \mathcal{A}_{\vartriangle}^{\hexagon} - 1,
    \\
    \text{(IIb):}& \quad (i,j) = (\widetilde{B},\widetilde{B}') \,\, \Rightarrow \,\, \mathcal{A}_{\triangledown}^{\triangledown} = \mathcal{A}_{\vartriangle}^{\hexagon}|_{l\rightarrow\tilde{l}},
    \\
    \text{(IIIb):}& \quad (i,j) = (\widetilde{A},\widetilde{B}') \,\, \Rightarrow \,\, \mathcal{A}_{\vartriangle}^{\triangledown} = \mathcal{A}_{\vartriangle}^{\hexagon},
    \\
    \text{(IVb):}& \quad (i,j) = (\widetilde{B},\widetilde{A}') \,\, \Rightarrow \,\, \mathcal{A}_{\triangledown}^{\vartriangle} = \mathcal{A}_{\triangledown}^{\hexagon}|_{l\rightarrow\tilde{l}} - 1.
\end{align}
\end{subequations}
Hence, all possible $C_{3z}$-$C_{3z}$ overlaps are avoided, and so we may conclude from the above analysis that only $C_{2z}$-$C_{2z}$ overlap will occur in this system. Hence, $AB$ stacked TBK will possesses a point group symmetry of $D_{2}$.

\section{RSP Overlaps: $\theta_{p} = 120^{\circ}$} \label{RSP_overlaps_120deg_sec}

Rather than twisting one layer about its isolated HSP with $C_{6z}$ symmetry, we may instead choose to rotate one layer about the bilayer HSP. In the case of bernal stacked graphene and interlocked Kagome this corresponds to the $AB$ site and up/down triangle overlap, respectively, which exhibit $D_{3d}$ symmetry. Applying the twist around this centre, we immediately guarantee the existence of a lattice of $C_{3z}$-$C_{3z}$ overlap points given that the twist origin may be labelled as a moir\'{e} lattice site. Hence, we need only check for $C_{6z}$-$C_{6z}$ overlap to determine whether the point group symmetry is $D_{3}$ or $D_{6}$. We note that, unlike the $\theta_{p} = 60^{\circ}$ scenario, we must consider both when $n$ is divisible by 3 and when $n$ is not divisible by 3.

\subsection{$n$ Divisible By 3}

Starting with $\partial_{p} f(\widetilde{\boldsymbol{\delta}}_{6'},\widetilde{\boldsymbol{\delta}}_{6};k,l,p,q;\theta_{c}) = 0$, we acquire
\begin{equation}
    \mathcal{A} = \frac{(2k+l)m^{2} + 2(2-3l)m\nu - 3(2k+l)\nu^{2}}{m^{2}+3\nu^{2}},
\end{equation}
which is not necessarily non-integer for $m,n,k,l \in \mathbb{Z}$. Let us therefore rearrange for $k$ and assume $\exists \, l, \mathcal{A} \in \mathbb{Z}$ s.t. $k \in \mathbb{Z}$,
\begin{equation}
    k = \frac{(\mathcal{A}-l)m^{2} + 2(3l-2)m\nu + 3(\mathcal{A}+l)\nu^{2}}{2(m^{2}-3\nu^{2})}.
    \label{120deg_k_n3}
\end{equation}
Substituting eq. \ref{120deg_k_n3} into the overlap function with $q = \mathcal{A} - 2p$ and setting the overlap function to zero yields
\begin{equation}
    l = -\mathcal{A} + 2p + \frac{2m(m(2+3\mathcal{A}-6p) - 3\mathcal{A}\nu)}{3(m^{2}+3\nu^{2})}.
\end{equation}
The final term of this expression cannot be integer as the numerator cannot be divided by 3. Therefore, by contradiction, $\nexists \, l \in \mathbb{Z}$ s.t. $k \in \mathbb{Z}$ and $f(\widetilde{\boldsymbol{\delta}}_{6'},\widetilde{\boldsymbol{\delta}}_{6};k,l,p,q;\theta_{c}) = 0$. Hence, when $\theta_{p} = 120^{\circ}$ and $\theta_{c}$ is associated to a choice of $n$ that is divisible by 3 bernal TBG and interlocked TBK will possess a $D_{3}$ point group symmetry.

\subsection{$n$ Not Divisible By 3}

As in section \ref{AB_TBK_guaranteed_RSP_overlap_sec}, we start by moving to the twist-symmetric frame and search for where the $C_{6z}$ RSPs of each layer lie on the $x$-axis. The RSP vectors in this frame are given by
\begin{subequations}
\begin{align}
    \widetilde{\mathbf{v}}_{6} &= R_{\frac{\theta_{c}}{2}} (\widetilde{\boldsymbol{\delta}}_{6} + k \mathbf{a}_{1} + l \mathbf{a}_{2}),
    \\
    \widetilde{\mathbf{v}}_{6'} &= R_{-\frac{\theta_{c}}{2}} (\widetilde{\boldsymbol{\delta}}_{6'} + p \mathbf{a}_{1} + q \mathbf{a}_{2}).
\end{align}
\end{subequations}
Setting the $y$ components of these vectors to zero yields,
\begin{subequations}
\begin{align}
    2n k + (3m+n) l &= 2m,
    \\
    2n p - (3m-n) q &= 2m.
\end{align}
\label{120deg_diophantine_eqs}
\end{subequations}
We now note that
\begin{equation}
    \text{GCD}(2n,(3m \pm n)) = \begin{cases}
        1, \quad \mathcal{P}_{m} \neq \mathcal{P}_{n} \\
        2, \quad \mathcal{P}_{m} = \mathcal{P}_{n} = -1 \\
        3, \quad n/3 \in \mathbb{Z}
    \end{cases},
\end{equation}
which will always divide $2m$ for the first two cases but never in the third cases, and so $k,l,p,q \in \mathbb{Z}$ satisfying these diophantine equations are ensured by B\'{e}zout's identity when $n$ is not divisible by 3. Given that $C_{6z}$ RSPs from each layer will lie on the $x$-axis, let us now discern if they are able to overlap. Solving eq. \ref{120deg_diophantine_eqs} for $l$ and $q$, we obtain
\begin{equation}
    l = \frac{2(m - nk)}{3m+n}, \qquad q = -\frac{2(m - np)}{3m-n},
    \label{120deg_lq_n_not_3}
\end{equation}
which we substitute into $\widetilde{\mathbf{v}}_{6} - \widetilde{\mathbf{v}}_{6'}$. Given the $y$ component of the resulting vector is always zero, we search for where its $x$ component vanishes to obtain the following relation between $k$ and $p$,
\begin{equation}
    k = \frac{(3m+n)p - 2m}{3m-n} = p + q.
    \label{120deg_k_n_not_3}
\end{equation}
Hence, if $p \in \mathbb{Z} \text{ s.t. } q \in \mathbb{Z}$, then $k \in \mathbb{Z}$. Now, let us check if this integer value of $k$ will also give $l \in \mathbb{Z}$. Substituting eq. \ref{120deg_k_n_not_3} into the expression for $l$ in eq. \ref{120deg_lq_n_not_3}, we find that $l = -q$, meaning that $l \in \mathbb{Z}$ when $p \in \mathbb{Z} \text{ s.t. } q \in \mathbb{Z}$. Therefore, $C_{6z}$-$C_{6z}$ overlap will always appear in bernal TBG and interlocked TBK when twisted about their HSP by a commensurate angle attributed to $n$ not being divisible by 3.

\subsection{Connection to $\theta_{p} = 60^{\circ}$}

When $n$ is divisble by 3 and $\theta_{p} = 120^{\circ}$, the rotational misalignment of the layers at an $AB$ (up/down) overlap site will be identical to the twisted bilayers with $n$ divisible by 3 and $\theta_{p} = 60^{\circ}$. Hence, the moir\'{e} pattern and point group for bernal TBG and interlocked TBK are equivalent. In contrast, when $n$ is not divisible by 3 and $\theta_{p} = 120^{\circ}$, the misalignment of overlapping hexagons cannot match that of the $AA$ system with $\theta_{p} = 60^{\circ}$ due to the unique labeling of each commensurate angle with $(m,n)$. However, given that commensurate angles are paired such that $\theta_{c}' = \pi/3 - \theta_{c}$ is associated to $n$ not being divisible by 3 when $\theta_{c}$ is given by a choice of $n$ that is divisible by 3 \cite{Scheer2022}, we can obtain the moir\'{e} pattern of one from the other provided the twist centre possesses $C_{6z}$ symmetry. Specifically (see the Appendix for details), by reflecting the $AA$ system in the $x$-axis and applying a global $\pi/3$ rotation, we arrive at the superlattice for a $\theta_{c}'$. After this transformation, the hexagon overlap for the $\theta_{p} = 60^{\circ}$ $AA$ stacked system will be identical to that of the bernal/interlocked system twisted about its $C_{3z}$ HSP.

\section{Conclusions}

In this paper considered the overlap of points with non-trivial rotational symmetry between in twisted bilayers of Kagome and graphene. If these rotational symmetry points do not overlap, the resulting moir\'{e} superlattice will not possess a non-trivial rotational symmetry. By constructing the untwisted bilayers in a high-symmetry stacking, the system will always possess a non-trivial rotational symmetry when one layer is rotated to a commensurate twist angle about the bilayer HSP or its own $C_{6z}$ HSP. When the twist is applied around a $C_{6z}$ centre for a high-symmetry stacking, the twist centre will possess a $C_{3z}$ or $C_{2z}$ point group for the interlocked and $AB$ Kagome bilayers. However, we have demonstrated that a higher symmetry point is guaranteed to exist at commensurate twist angles.

In particular, we shown that RSPs with $C_{jz}$ matching the rotational symmetry of the untwisted bilayer can always be identified. Consequently, in the twist-symmetric frame, the $x$-axis corresponds to the in-plane axis with $C_{2}'$ symmetry, thus preserving the dihedral symmetry of the bilayer. However, the mirror planes of the individual layers will not be aligned, preventing the twisted bilayer from hosting a higher point group symmetry. We therefore conclude that the point group symmetry for TBG and TBK will be $D_{j}$ with $j$ set by the rotational symmetry of the untwisted bilayer.

Finally, we recover the results of Mele \cite{Mele2010} when the twist is applied about the HSP of the interlocked Kagome bilayer using the the overlap function defined in section \ref{Overlap_function_sec}. The moir\'{e} patterns obtained from this alternative twist centre can be mapped onto those generated by the $C_{6z}$ twist centre. The results of this paper will help to inform future studies on twisted Kagome bilayers as systems hosting the monolayer lattice continue to be discovered and studied in the context of band engineering.

\begin{acknowledgements}
    DTSP acknowledges funding from the UK EPSRC grants EP/X012557/1 and EP/T034351/1. DTSP would also like to thank Prof. Joseph J. Betouras for useful discussions about the method presented in this paper, as well as Prof. Aires Ferreira and Dr Mark T. Greenaway for their comments on this manuscript.
\end{acknowledgements}

\section*{Appendix: Two Classes of Commensurate Angles} \label{Appendix_sec}

We start by defining the two lattices constituting a bilayer prior to twisting as $L_{1}$ and $L_{2}$ for the bottom and top layers, respectively. The top layer will displaced relative to the first by $\mathbf{d}$ to form a high-symmetry stacking and the origin will be located at the centre of a hexagon for $L_{1}$. The lattices are thus given by
\begin{equation}
\begin{split}
    L_{1} &= \{ \mathbf{R}_{i} + \boldsymbol{\tau}_{\alpha} \, | \, \mathbf{R}_{i} \in L, \alpha \in S \},
    \\
    L_{2} &= \{ \mathbf{R}_{i} + \boldsymbol{\tau}_{\alpha} + \mathbf{d} \, | \, \mathbf{R}_{i} \in L, \alpha \in S \},
\end{split}
\end{equation}
where $L$ is the underlying hexagonal lattice independent of sublattice geometry, $S$ is the sublattice geometry with $S = \{A,B\}$ ($S = \{A,B,C\}$) for graphene (Kagome), and $\boldsymbol{\tau}_{\alpha}$ is the position of sublattice $\alpha$ in the monolayer unit cell. We take $\boldsymbol{\tau}_{\alpha}$ such that $L_{1}$ possesses mirror reflection symmetry in the $x$-axis, $M_{x}$ (i.e. $y \rightarrow -y$), see Fig. \ref{Stacking_orders}. For all stacking configurations of the graphene and Kagome bilayers, the second layer under this mirror reflection can be reinterpreted as a rotation by an angle of $-60^{\circ}$, $L_{2}' = M_{x} L_{2} = R_{-\frac{\pi}{3}} L_{2}$.

Let us now consider rotating $L_{1}$ by an arbitrary angle of $\pm\theta$, $L_{1}^{\theta} = R_{\theta} L_{1}$ and $L_{1}^{\bar{\theta}} = R_{-\theta} L_{1}$. The two twisted bilayers of concern are now $\mathcal{L}^{\theta} = L_{1}^{\theta} \cup L_{2}$ and $\mathcal{L}^{\bar{\theta}} = L_{1}^{\bar{\theta}} \cup L_{2}$. Courtesy of $M_{x}L_{1} = L_{1}$, we have $M_{x}L_{1}^{\bar{\theta}} = L_{1}^{\theta}$, and so $M_{x}\mathcal{L}^{\bar{\theta}} = L_{1}^{\theta} \cup L_{2}'$. Moreover, because $R_{\frac{\pi}{3}}L_{1}^{\theta} = L_{1}^{\theta}$, we see that $R_{\frac{\pi}{3}}M_{x}\mathcal{L}^{\bar{\theta}} = \mathcal{L}^{\theta}$. Since mirror reflection and global rotation preserve the overlap of RSPs, we see that the point group symmetry for a twist of $-\theta$ must be equivalent to that of $\theta$. Finally, having applied the twist about the $C_{6z}$ symmetry point of the bottom layer, a twist of $-\theta$ is equivalent to $\pi/3 - \theta$. Therefore, the set of $\theta_{c}$ with 3 not dividing $n$ will have the same point group symmetry as those with a choice of $n$ that is divisible by 3 when $\theta_{p} = 60^{\circ}$.

%

\end{document}